# Enhanced thermal sensitivity of MEMS bolometers integrated with two-dimensional phononic crystals

Ya Zhang[1,2,a)], Boqi Qiu[1], Naomi Nagai[1], Masahiro Nomura[1,3,4], Sebastian Volz[4], and Kazuhiko Hirakawa[1,3,4,b)]

[1] *Institute of Industrial Science, University of Tokyo, 4-6-1 Komaba, Meguro-ku, Tokyo 153-8505, Japan*

[2] *Institute of Engineering, Tokyo University of Agriculture and Technology, Koganei, Tokyo, 184-8588, Japan*

[3] *Institute for Nano Quantum Information Electronics, University of Tokyo, 4-6-1 Komaba, Meguro-ku, Tokyo 153-8505, Japan*

[4] *LIMMS/CNRS-IIS, UMI 2820, 4-6-1 Komaba, Meguro-ku, Tokyo 153-8505, Japan*

We have fabricated two-dimensional phononic crystal (PnC) structures on GaAs doubly-clamped microelectromechanical system (MEMS) beam resonators to modulate their thermal properties.  Owing to the reduction in the thermal conductance of the MEMS beams by introducing the PnC structures, the MEMS bolometers with the PnC structures show 2-3 times larger thermal sensitivities than the unpatterned reference sample.  Furthermore, since the heat capacitance of the MEMS beams is also reduced by introducing the PnCs, the thermal decay time of the patterned MEMS beams is increased only by about 30-40 %, demonstrating the effectiveness of the PnCs for enhancing the thermal sensitivities of bolometers without significantly deteriorating their operation bandwidths.

―――――――――――

a) Electronic mail: zhangya@go.tuat.ac.jp

b) Electronic mail: hirakawa@iis.u-tokyo.ac.jp

Micro-electromechanical system (MEMS) resonators[1-3] are very attractive for sensing applications. Owing to their high quality (Q)-factors, the MEMS resonators can detect a small change in the resonance frequencies and can be used to detect changes in mass,[4-8] charge,[9,10] spin orientation,[11,12] and temperature[13,14]. Recently, we have developed an uncooled, sensitive and fast bolometer by using a doubly clamped GaAs MEMS beam resonator for terahertz (THz) sensing applications.[15-17] The MEMS resonator detects THz radiation by measuring the shift in the resonance frequency caused by heating of the MEMS beam.[15-17] Since the responsivity of the MEMS resonators is inversely proportional to the thermal conductance of the MEMS beam, $G_T$, it is preferable to decrease $G_T$ for enhancing the responsivity. On the other hand, the thermal time constant $\tau_D$ (= $C_T/G_T$; $C_T$ is the heat capacitance of the MEMS beam) increases when $G_T$ is decreased, leading to the reduction in the detection speed. This trade-off between the responsivity and the detection bandwidth exists in all kinds of thermal sensors.

The phononic crystals (PnCs)[18,19] are the structures that have a periodic modulation in the elastic modulus and/or mass density. The PnC structures such as slabs with one-dimensional (1D)[20] or two-dimensional (2D)[21-23] hole arrays have been proposed to engineer the thermal properties of materials. The hole-array-based PnC structures are promising for improving thermal responsivities of the MEMS resonators; the PnC structure can reduce the thermal conductance, $G_T$, of the MEMS beam by decreasing the cross section of the beam. Furthermore, the PnC structure reduces the heat capacitance of the beam, $C_T$, by decreasing the material volume. Therefore, the increase in $\tau_D$ due to the decrease in $G_T$ is partly compensated by the reduction in $C_T$ and it is possible to enhance the thermal responsivity, while keeping a fast detection speed of the MEMS thermal sensors.

In this work, we have fabricated 2D PnC structures on the MEMS beam resonators to modulate their thermal properties. Homogenous hole arrays of square lattices were formed on the GaAs MEMS beams. Owing to the reduction in the thermal conductance of the MEMS beams by introducing the PnC structures, the samples with the PnC structures show 2-3 times larger thermal sensitivities than a reference unpatterned sample. Furthermore, because the heat capacitance of the MEMS beams is also reduced by the PnC structures, the thermal decay time of the samples with the PnC structures is increased only by about 30%~40%, demonstrating the effectiveness of the PnCs for enhancing the thermal responsivities of bolometers without deteriorating their operation bandwidths significantly.

The wafer used for fabricating the doubly clamped MEMS beam resonators was grown by molecular beam epitaxy.[24] After growing a 200-nm-thick GaAs buffer layer and a 3-μm-thick $Al_{0.7}Ga_{0.3}As$ sacrificial layer on a (100)-oriented semi-insulating GaAs substrate, the beam layer was formed by depositing a 50-nm-thick GaAs layer, a $GaAs/Al_{0.3}Ga_{0.7}As$ superlattice structure, and a 400-nm-thick GaAs layer. We subsequently grew a 20-nm-thick Si-doped GaAs layer, a 70-nm-thick $Al_{0.3}Ga_{0.7}As$ layer and a 10-nm-thick GaAs capping layer. The fabrication process for the MEMS resonators with PnC structures were schematically shown in Fig. 1(a). The PnC structures of the square lattice were patterned on the beam by using electron-beam lithography. The holes were formed by using reactive ion etching with $Cl_2$ gas and a rf power of 200 W at 50 °C for 80 s. The suspended beam structure was formed by selectively etching the sacrificial layer with diluted hydrofluoric acid[24]. Figure 1(b) shows an optical microscope image of a fabricated MEMS beam resonator (100×30×0.6 μm³) with a 2D PnC structure of a hole diameter $d$ = 500 nm and the neck size (the distance between neighboring holes) $n$ = 500 nm. The inset of Fig. 1(b) shows a blow-up of an SEM image of the PnC structure, showing that the fabricated hole array is homogeneous. We fabricated MEMS beams with the PnC structures of various sizes, *i. e.*, $d/n$ = 500 nm/500 nm, 500 nm/400 nm, 500 nm/300 nm, 300 nm/300 nm, and 300 nm/200 nm. In addition, we fabricated a reference sample without a PnC structure. The Si-doped GaAs layer and the top metal gates (15-nm-thick NiCr) on the two ends of the MEMS beam form two piezoelectric capacitors. An *ac* voltage was applied to one of the piezoelectric capacitors to drive the beam and the induced resonant beam motion was monitored by a laser Doppler vibrometer, as schematically shown in Fig.1(c). The resonance signal is input to a phase locked loop (PLL) to provide a feedback control for maintaining a self-oscillation, as we reported elsewhere[17]. On the MEMS beam, we deposited a 15-nm-thick NiCr layer as a heater for calibrating the responsivity of the MEMS resonator, whose sheet resistance was ~500 Ω/□. All the measurements were performed in a vacuum (~$10^{-4}$ torr) at room temperature.

Figure 2(a) shows the measured oscillation spectra of a MEMS resonator with a PnC structure ($d/n$ = 300 nm/200 nm) at various driving voltages ($V_D$ = 10-80 mV) applied to the piezoelectric capacitor. The resonance frequency, $f_0$, was about 213.8 kHz and the Q-factor was about 1,500. $f_0$ was higher than that of the reference MEMS resonator (~167.8 kHz). This is because the etching of the sacrificial layer develops through the PnC holes for the PnC samples, whereas the etching proceeds only from the sides of the beam for the reference sample. This difference in the etching paths makes a difference in the length of the etching undercut, resulting in a shorter effective beam length for the PnC samples. When $V_D$ exceeds 50 mV, the MEMS resonator shows a nonlinear hardening effect.

When an input power to the NiCr film, $P_{in}$, is increased from 0 to 50 µW, $f_0$ is reduced due to the thermal stress of the beam. Figure 2(b) shows the normalized frequency shift, $\Delta f/f_0$, as a function of the input heating power, $P_{in}$, for two MEMS resonators with PnC structures ($d/n$ = 300 nm/300 nm and 300 nm/200 nm) and a reference MEMS resonator without the PnC. From the slope of the frequency shift shown in Fig. 2(b), we determined the thermal responsivity, $R \equiv \Delta f/f_0 P_{in}$, for the samples. $R$ is increased from ~393 W$^{-1}$ for the reference sample to ~712 W$^{-1}$ for $d/n$ = 300 nm/300 nm and ~892 W$^{-1}$ for $d/n$ = 300 nm/200 nm, indicating that the PnC is effective in increasing the thermal responsivity of the MEMS detectors.

To back up our interpretation, we have calculated the thermal conductance of the unit cell of the square PnC lattice, $g_T$. Here, we introduce the porosity of the beam, $p$, which is defined by the ratio of the material volume removed from the beam to the volume of the beam before fabricating PnC structures[21],

$$p = \frac{pd^2}{4(d+n)^2}.\qquad(1)$$

As seen in Eq. (1), $p$ is determined only by the ratio $n/d$. In the calculation, we assumed that the left and right boundaries of the PnC unit cell has a temperature difference, $\Delta T$, and we calculated the heat power, $P_{\text{heat}}$, transmitted through the unit cell. $g_T$ is defined as $g_T \equiv P_{\text{heat}}/\Delta T$. The red dashed line in Fig. 3 shows $g_T$ as a function of the porosity of the PnCs. The PnC structures fabricated in this work have a porosity of 0.2-0.3 and we expect that $g_T$ is reduced by 40%~50% in our samples. In the figure, the inverse of the thermal responsivities of the PnC samples normalized by that of the unpatterned sample are also plotted. The measured inverse responsivities are in good agreement with the calculated $g_T$, indicating that the enhancement in $R$ is indeed due to the reduction in $G_T$ by introducing the PnCs.

Next, to examine the effect of the PnC structures on the detection speed of the MEMS thermal sensors, we measured the thermal decay time, $\tau_D$, of the beam by measuring the heat signal as a function of the modulation frequency, $f_m$. We drove the MEMS beam resonator in a self-oscillation mode by using a PLL and applied an ac voltage to the NiCr heater to generate a modulated heat of ~2.3 µW on the beam. From a simple thermal decay theory, the frequency shift, $\Delta f$, and the thermal decay time, $\tau_D$, have a relationship expressed by;

$$\Delta f(f_m) = \frac{\Delta f_0}{\sqrt{1+(2\pi \tau_D f_m)^2}} G_{PLL}(f_m) \quad , \tag{2}$$

where $\Delta f_0$ is the frequency shift when the heat modulation frequency $f_m = 0$. $G_{\text{PLL}}(f_m)$ expresses the circuit response of the PLL.

We first characterized $G_{\text{PLL}}(f_m)$ to calibrate the effect of the demodulation bandwidth (BW) of the PLL. We input a frequency-modulated (FM) ac signal (amplitude = 1V, carrier frequency = 200 kHz, and FM depth 1 kHz) to the PLL to simulate the signal from the MEMS bolometer.[17] We swept $f_m$ and measured the demodulated output of the PLL to obtain $G_{\text{PLL}}(f_m)$. Then, we can obtain the intrinsic frequency response of the beams as $\Delta f(f_m)/G_{\text{PLL}}(f_m)$, which is only determined by the thermal decay process in the MEMS beam. Figure 4(a) plots $\Delta f(f_m)/G_{\text{PLL}}(f_m)$ for a reference MEMS resonator without the PnC and typical two MEMS resonators with PnC structures ($d/n$ = 300 nm/300 nm and 300 nm/200 nm). As seen in the figure, the signals for the PnC samples decrease slightly faster than that of the reference sample, indicating that the thermal decay times, $\tau_D$, are slightly increased by introducing the PnC structures.

By using numerical fitting of Eq. (2) to $\Delta f(f_m)/G_{PLL}(f_m)$, we obtained $\tau_D$ for the MEMS beams with the PnCs and that for the reference MEMS beam. Figure 4(b) plots $\tau_D$ of the MEMS beams with PnC structures as a function of $p$. $\tau_D$ is increased from ~75.8 μs for the reference sample to ~97.4 μs for $d/n$ = 300 nm/300 nm and ~112.5 μs for $d/n$ = 300 nm/200 nm. From Fig. 4(b), we see that the increase in $\tau_D$ by introducing the PnC structures is typically 30-40%. Compared with the improvement in responsivity (2-3 times), the reduction in the thermal BW is smaller, demonstrating effectiveness of the PnCs for partly resolving the trade-off between the responsivity and the bandwidth of the thermal sensors.

Finally, we have characterized the noise equivalent power (NEP) of the MEMS beam resonators. We drove the MEMS beam resonators in a self-oscillation mode by using a PLL with a demodulation band width of 1 kHz and measured the frequency noise spectra, $n_f$. Figure 5 plots $n_f$ as a function of $f_m$ for two MEMS resonators with PnCs and a reference MEMS resonator. The samples with PnCs have slightly smaller frequency noise than the reference sample. Since the NEP of the MEMS resonator is expressed as NEP $\equiv n_f/f_0 R$, the NEP is reduced by the enhanced thermal responsivities for the PnC samples. The minimum NEP for the reference sample was ~110 pW/Hz$^{0.5}$ at $f_m$ = 1 kHz, whereas the NEP for the sample with $d/n$ = 300 nm/300 nm is ~50 pW/Hz$^{0.5}$ and that for the sample with $d/n$ = 300 nm/200 nm is ~40 pW/Hz$^{0.5}$. Since the detector sensitivity is defined by 1/NEP, the MEMS resonators with the PnCs have 2-3 times larger sensitivities than the unpatterned reference sample.

In summary, we have fabricated two-dimensional PnC structures on GaAs doubly-clamped MEMS beam resonators to modulate their thermal properties. Owing to the reduction in the thermal conductance of the MEMS beams by introducing the PnC structures, the samples with PnC structures show 2-3 times larger thermal responsivities than an unpatterned reference sample. Furthermore, since the heat capacitance of the MEMS beams is also reduced by introducing the PnC structures, the thermal decay time of the samples with the PnC structures is increased only by about 30%~40%, demonstrating the effectiveness of the PnCs for enhancing the thermal sensitivities of bolometers without deteriorating their operation bandwidths very much.

We thank Y. Watanabe for his contribution at the early stage of this work. This work has been supported by JST Collaborative Research Based on Industrial Demand (JPMJSK1514), KAKENHI from JSPS (17K14654), and MEXT Grant-in-Aid for Scientific Research on Innovative Areas "Science of hybrid quantum systems" (15H05868).

**Figure Captions**

Fig. 1 (color online) (a) Fabrication processes of a MEMS resonator with a PnC structure. (b) Microscope image of a fabricated GaAs MEMS beam resonator ($100\times30\times0.6$ $\mu m^3$) with a 2D PnC structure of a hole diameter $d = 500$ nm and a neck size $n = 500$ nm. The Si-doped GaAs layer and the top gates on both ends of the beam form two piezoelectric capacitors, $C_1$ and $C_2$. A 15-nm-thick NiCr THz absorbing layer was deposited on the beam. This metal film was used also as a heater to calibrate the thermal responsivity of the resonator. The inset shows a blow-up of an SEM image of the PnC structure. (c) Schematic illustration for the measurement setup. An *ac* voltage was applied to one of the piezoelectric capacitors to drive the beam and the induced beam motion was monitored by a laser Doppler vibrometer. The motion signal is input to a phase locked loop (PLL) to provide a feedback control for maintaining the self-oscillation.

Fig. 2 (color online) (a) Resonance spectra of a MEMS beam resonator with a PnC structure (d/n = 300nm/200nm) measured by using an open-loop circuit (lock-in amplifier) at various driving voltages ($V_D = 10\text{-}80$ mV). (b) The normalized frequency shift ($\Delta f/f_0$) as a function of the input heating power, $P_{in} = 0\text{-}50$ $\mu$W, for two MEMS resonators with PnC structures (d/n = 300nm/300nm and 300nm/200 nm) and that for a reference MEMS resonator. The numbers in the figure are the thermal responsivities $R \equiv \Delta f/f_0 P_{in}$ of these three samples determined from the slope of $\Delta f/f_0$.

Fig. 3 (color online) Red dashed line plots the normalized thermal conductance of the unit cell of the square PnC lattice calculated as a function of the porosity of the PnC structure for a dimensionless unit cell (see the inset) by using finite element method. The symbols show the normalized thermal conductance of the PnC MEMS beams with various $d$ and $n$, derived from the measured thermal decay time of the beams shown in Fig. 4(b), as a function of the porosity of the PnC structure.

Fig. 4 (color online) (a) $\Delta f(f_m)/G_{PLL}$ measured for two PnC MEMS resonators (d/n=300nm/300nm and 300nm/200nm) and a reference sample. (b) Thermal decay time $\tau_D$'s of the beams with various PnC structures are plotted as a function of the porosity of the PnC structure. $\tau_D$'s were derived from the $\Delta f(f_m)/G_{PLL}$ spectra.

Fig. 5 (color online) Frequency noise spectra of the PnC MEMS resonators (d/n = 300nm/300nm and 300nm/200 nm) and the reference sample. The spectra were measured when the MEMS resonators were driven in a self-oscillation mode with a resonance amplitude of ~150 nm by a PLL with a modulation bandwidth of 1 kHz.

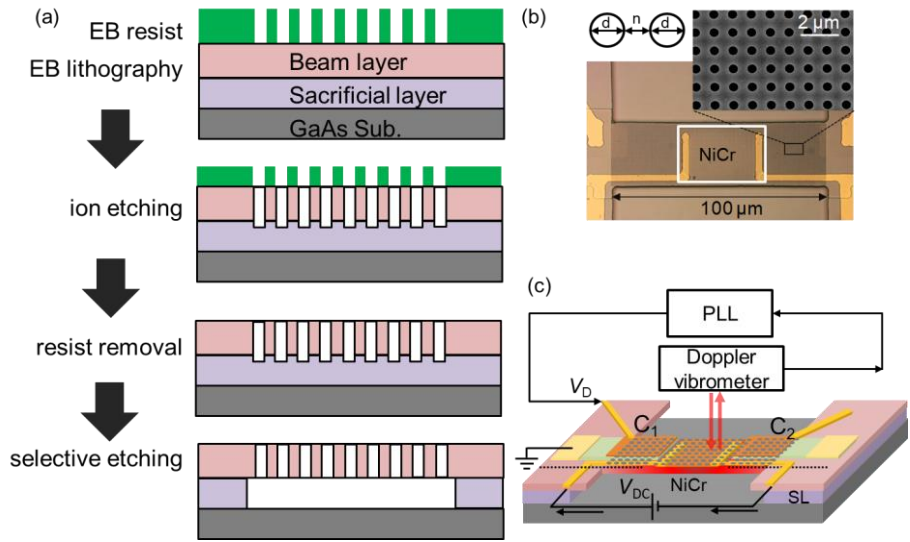

Fig. 1 Y. Zhang, et al.

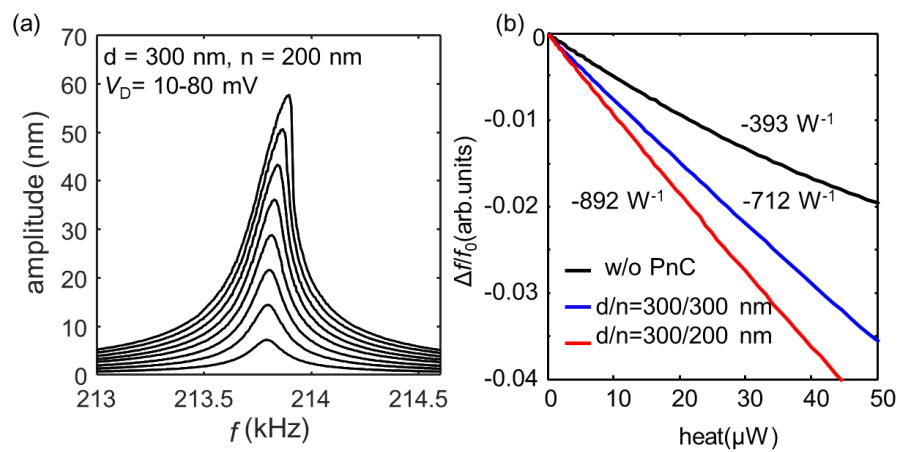

Fig. 2 Y. Zhang, et al.

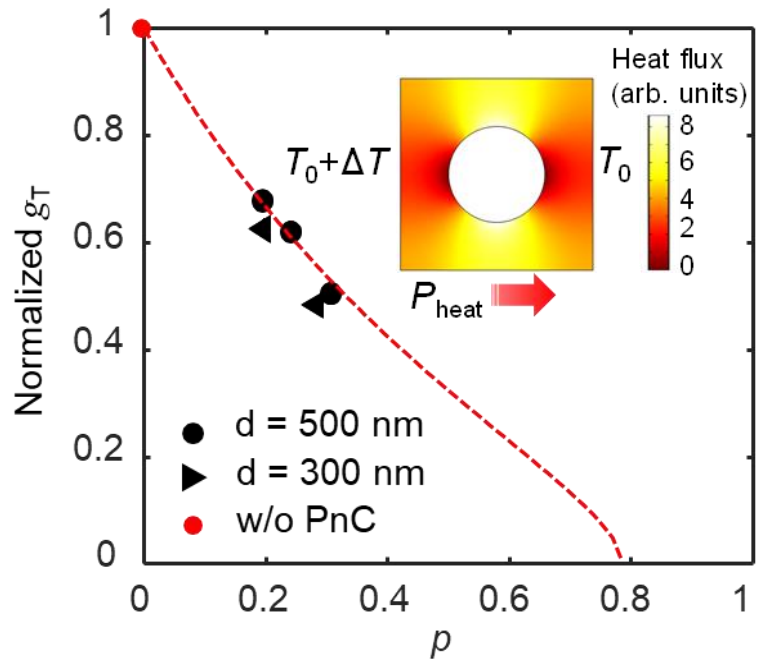

Fig. 3 Y. Zhang, et al.

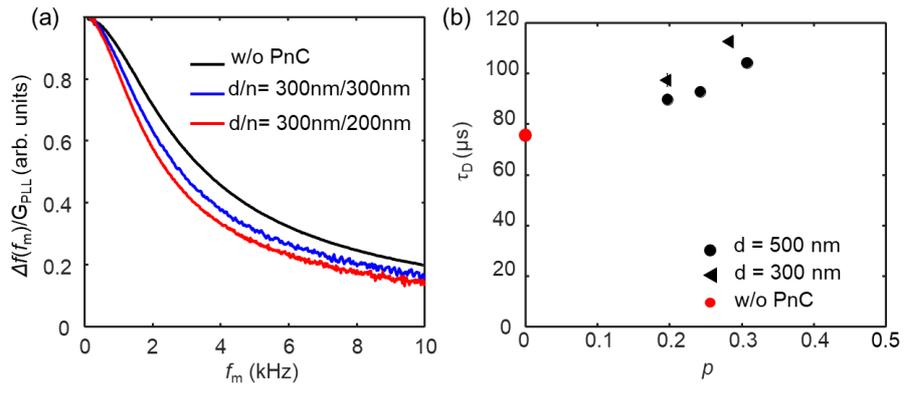

Fig. 4 Y. Zhang, et al.

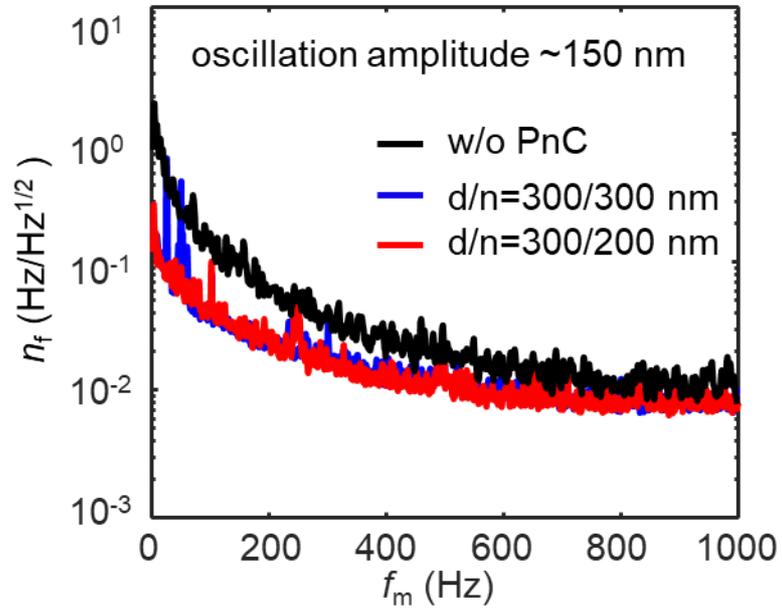

Fig. 5  Y. Zhang, et al.